\documentclass[final,5p,times,twocolumn]{elsarticle}

\usepackage{graphicx}

\usepackage{amssymb}
\usepackage{amsmath}

\usepackage{url}

\journal{Nuclear Instruments and Methods in Physics Research Section A}

\begin{document}

\begin{frontmatter}



\title{Simulation Chain for Acoustic Ultra-high Energy Neutrino Detectors}
\author{M.~Neff\corref{cor1}}
\ead{max.neff@physik.uni-erlangen.de}
\ead[url]{http://www.ecap.nat.uni-erlangen.de/acoustics/}
\cortext[cor1]{Corresponding author}
\author{G.~Anton}
\author{A.~Enzenh\"ofer}
\author{K.~Graf}
\author{J.~H\"o\ss l}
\author{U.~Katz}
\author{R.~Lahmann}


\address{Friedrich-Alexander-Universit\"{a}t Erlangen-N\"{u}rnberg, Erlangen Centre for Astroparticle Physics, Erwin-Rommel-Str. 1, 91058 Erlangen, Germany}

\begin{abstract}
Acoustic neutrino detection is a promising approach for large-scale ultra-high energy neutrino detectors in water. In this article, a Monte Carlo simulation chain for acoustic neutrino detection devices in water is presented. It is designed within the SeaTray/IceTray software framework. Its modular architecture is highly flexible and makes it easy to adapt to different environmental conditions, detector geometries, and hardware. The simulation chain covers the generation of the acoustic pulse produced by a neutrino interaction and the propagation to the sensors within the detector. In this phase of the development, ambient and transient noise models for the Mediterranean Sea and simulations of the data acquisition hardware, similar to the one used in ANTARES/AMADEUS, are implemented. A pre-selection scheme for neutrino-like signals based on matched filtering is employed, as it can be used for on-line filtering. To simulate the whole processing chain for experimental data, signal classification and acoustic source reconstruction algorithms are integrated. In this contribution, an overview of the design and capabilities of the simulation chain will be given, and some applications and preliminary studies will be presented.
\end{abstract}


\end{frontmatter}


\section{Introduction}
\label{sec1:intro}
Acoustic neutrino detection uses the effect that ultra-high energy (UHE) neutrinos can produce a detectable acoustic pulse according to the thermo-acoustic model\,\cite{Askariyan1979267}. This model describes the generation of an acoustic pulse due to the local heating of the medium by a hadronic shower, which is caused by an UHE neutrino interaction. The fast deposition and slow dissipation of the energy by the cascade in the medium leads to a bipolar pulse (BIP) and due to the cylindrical geometry of the shower the wave propagates through the medium in a disk-like shape perpendicular to the main axis of the cascade. Given the expected low flux of neutrinos with energies in excess of 100\,PeV, a potential acoustic neutrino telescope must have large dimensions of presumably $\gtrsim$ 100\,km$^3$. The attenuation length of sound in water is of the order of 1\,km for the peak spectral density of around 10\,kHz, allowing for a less dense instrumentation. In this article, the efforts taken to develop a complete simulation chain to reproduce the acoustic pulse generation, the detector properties, and the deep-sea acoustic environment are reported.
\section{Simulation Chain}
\label{sec2:sim_chain}
\begin{figure}[tb]
\centering
\includegraphics[width=0.4\textwidth]{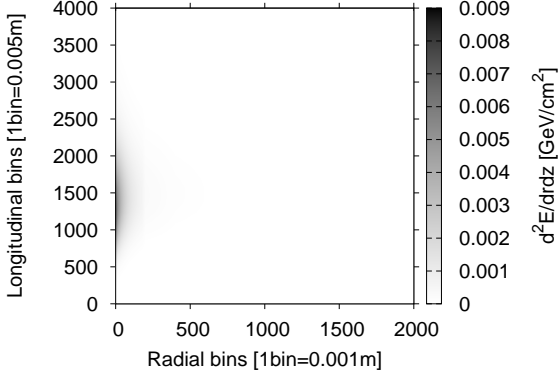}
\caption{
The longitudinal and radial distribution of the deposited energy of the shower in water. The total shower energy is $10^{11}$\,GeV.
\label{distribution}}
\end{figure}

\begin{figure}[tb]
\centering
\includegraphics[width=0.4\textwidth]{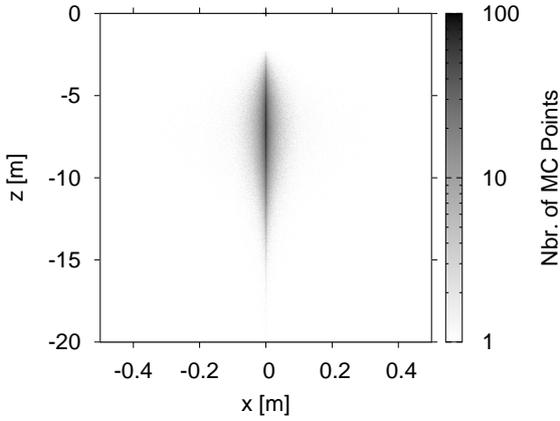}
\caption{
MC point distribution generated proportional to the deposited energy distribution of a $10^{11}$\,GeV hadronic shower as in Fig.~\ref{distribution}. Bin sizes are 0.001\,m in $x$ and 0.005\,m in $z$, where the points have been projected from a three-dimensional distribution upon the $xz$-plane.
\label{shower}}
\end{figure}

\begin{figure}[tb]
\centering
\includegraphics[width=0.35\textwidth]{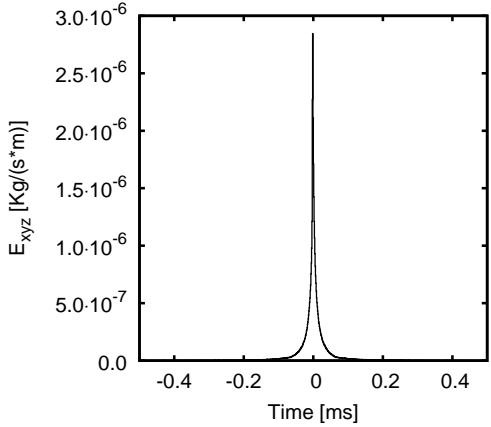}
\caption{
The velocity potential $E_{xyz}(t)$ as a function of the signal propagation time, which is normalized to the mean propagation time. The total energy deposition of the shower is $10^{11}$\,GeV at a distance of 1000\,m from the sensor.
\label{velopot}}
\end{figure}

\begin{figure}[tb]
\centering
\includegraphics[width=0.35\textwidth]{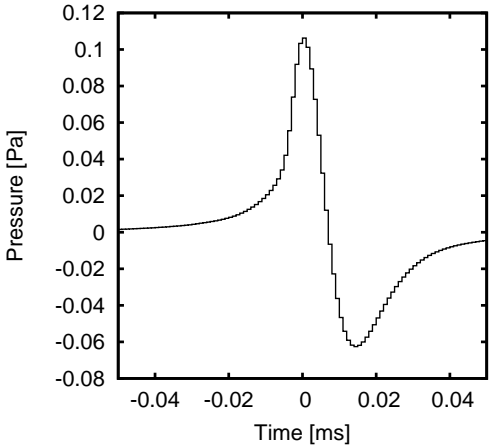}
\caption{
Simulated acoustic pressure signal for a $10^{11}$\,GeV shower as recorded at the position of the sensor. The distances between shower and sensor is 1000\,m. The bipolar shape of waveform with an asymmetric tail is recognisable.
\label{bip}}
\end{figure}

\begin{figure}[tb]
\centering
\includegraphics[width=0.4\textwidth]{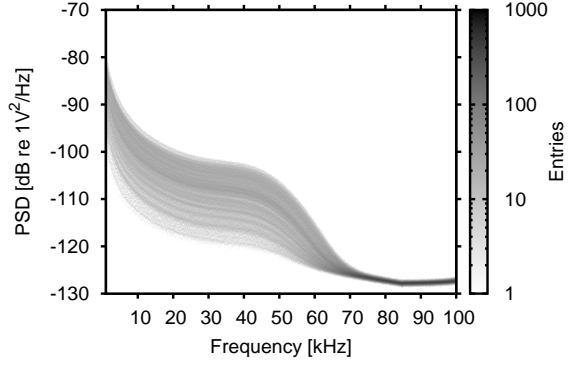}
\caption{
Power spectrum density (PSD) of the ambient noise as produced by the model described in the text. This scatter plot shows the PSD for the different levels of the weather conditions and shipping activities.
\label{noise_psd}}
\end{figure}

\begin{figure}[tb]
\centering
\includegraphics[width=0.4\textwidth]{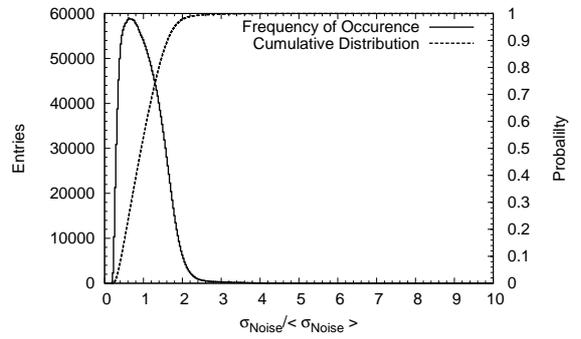}
\caption{
Frequency of occurrence distribution of the ambient noise level as function of the ratio between the noise level $\sigma_{noise}$ of the sample and the mean noise level $\langle\sigma_{noise}\rangle$. This simulation is based on an ambient noise model as described in the text. The left-side y-axis shows the number of occurrences and the right-side y-axis the probability that a noise level occurs.
\label{noise_occurrence}}
\end{figure}

In this section, the design of the simulation chain and the capabilities of the different modules will be presented. The simulation chain consists of the following stages, which build up on each other to create a simulated event:
\begin{itemize}
\itemsep 0pt
\item{An interaction vertex is located at a random position in a given volume around the detector and the energy and direction of the shower are set randomly.}
\item{The formation of the shower and the resulting acoustic signal, which is generated by an UHE neutrino interaction, are simulated.}
\item{The acoustic environment of the deep sea is reproduced including both the ambient and transient noise conditions.}
\item{The data acquisition (DAQ) hardware is simulated including the system response and inherent noise of the sensors and read-out electronics.}
\item{A pre-selection scheme is applied.}
\end{itemize}
The simulation chain was designed within the SeaTray/IceTray software framework\,\cite{Claudio2009107, DeYoung:865626}. Its modular architecture is highly flexible and makes it easy to adapt it to different environmental conditions, data acquisition hardware, and detector geometries. The generation of the Monte Carlo (MC) shower is the first step described here. The MC shower is produced from a parameterization, which is based on work by the ACoRNE collaboration\,\cite{S.Bevan:2007wd, Bevan:2009rr}. This parameterization describes the distribution of the deposited energy in the surrounding medium, in this case water. The longitudinal and radial energy density distribution of a $10^{11}$\,GeV shower is shown in Fig.~\ref{distribution}. From this energy distribution, a point distribution is produced with a point density proportional to the energy density distribution. The representation of the shower is shown in Fig.~\ref{shower}; it is about 20\,m long and has a diameter of about 10\,cm. After the cascade has been simulated, the acoustic pulse and its propagation to the sensors within the detector are calculated. The deposited energy of the shower produces a local heating of the medium. With respect to hydrodynamical time scales, the energy deposition at time $t_0$ is instantaneous and the dissipation of the energy is slow in comparison. The energy deposition $\epsilon (\mathbf{r},t)$ can be factorized into a spatial and temporal part using the Heaviside function.
\begin{equation}
\epsilon (\mathbf{r},t) = \tilde{\epsilon} (\mathbf{r})\Theta (t-t_0) \; \Rightarrow \;
\frac{\partial }{\partial t} \epsilon(\mathbf{r},t) = \tilde{\epsilon} (\mathbf{r})\delta (t-t_0)
\label{eq:epsilon}
\end{equation}
Assuming a total energy deposition $E$ with cylindrical symmetry, the spatial part $\tilde{\epsilon} (\mathbf{r})$ can be expressed for the longitudinal and radial positions $z$ and $r$ in the shower as:
\begin{equation}
\tilde{\epsilon} (\mathbf{r}) = \frac{1}{E} \frac{1}{2\pi r} \frac{d^2E}{drdz}
\label{eq:epsilon_part}
\end{equation}
Inserting Eq.~\ref{eq:epsilon} and Eq.~\ref{eq:epsilon_part} into the expression for the pressure p, following\,\cite{PhysRevD.19.3293}:
\begin{equation}
p(\mathbf{r},t) = \frac{\alpha}{4\pi C_p}\int {\frac{d^3 r'}{|\mathbf{r} - \mathbf{r'}|} \frac{\partial^2}{\partial t^2}\epsilon(\mathbf{r'},t-\frac{|\mathbf{r} - \mathbf{r'}|}{c_s})},
\label{eq:pressure_int}
\end{equation}
where $\alpha$ is the thermal expansion coefficient, $C_p$ is the specific heat capacity at constant pressure and $c_s$ the speed of sound in the medium. Eq.~\ref{eq:pressure_int} can be further reduced to:
\begin{equation}
p(\mathbf{r},t) = \frac{E\alpha}{4\pi C_p}\int \frac{d^3 r'}{R} \tilde{\epsilon}(\mathbf{r'}) \frac{d}{dt}\delta(t-\frac{R}{c_s}),
\label{eq:pressure_int_mod}
\end{equation}
where $R=|\mathbf{r} - \mathbf{r'}|$ is the distance between the shower maximum and the sensor. A velocity potential can be defined:
\begin{equation}
E_{xyz}(t) = \frac{E\alpha}{4\pi C_p}\int \frac{d^3 r'}{R} \tilde{\epsilon}(\mathbf{r'}) \delta(t-\frac{R}{c_s}).
\label{eq:velopot}
\end{equation}
As described before, the distribution of points within the MC shower is proportional to the energy density distribution of the shower. So the pressure at a sensor in a distance $R$ from the shower can be numerically calculated. The signal propagation time from each point within the MC shower to a sensor in the detector is calculated and entered into a histogram with a bin-width according to the sampling rate (here $1\,\mu s$) and a size big enough to hold the distribution (here $2^{15}$ bins). After normalising each bin with the number of points in the shower, scaling it with the contant term (see Eq.~\ref{eq:velopot}), and dividing by the mean distance to the shower, this results in the velocity potential $E_{xyz}(t)$ as shown in Fig.~\ref{velopot}. The Fourier Transform of Eq.~\ref{eq:pressure_int_mod} including Eq.~\ref{eq:velopot} can be written as:
\begin{equation}
\hat{p}(\omega) = \int \frac{d}{dt} E_{xyz}(t) e^{-i\omega t} dt 
= i\omega \hat{E}_{xyz}(\omega),
\label{eq:pressure_int_fft}
\end{equation}
taking into account the basic property of the Fourier Transform that the derivative in the time domain is the same as multiplying by $i\omega$ in the frequency domain. $\hat{E}_{xyz}(\omega)$ is derived form the histogram $E_{xyz}(t)$ using the Fast Fourier Transform (FFT). The frequency dependent sound attenuation in sea water is multiplied with the pressure signal in the frequency domain. The attenuation is based on a model by Ainslie and McColm\,\cite{1998ASAJ..103.1671A} extended to its complex representation. This procedure leads to the characteristic acoustic signal \--- a bipolar pulse \--- at the sensor as shown in Fig.~\ref{bip}.\\
The background for acoustic neutrino detection in the deep sea consists of two different types of noise: transient and ambient noise. Transient noise signals are short, but of high amplitude, and can mimic bipolar pulses from neutrino interactions. In the simulation, four types of transient signals are implemented so far: bipolar and multipolar pulses, sinusoidal signals, and signals with brown noise frequency characteristics. Sources of these four types can be anthropogenic sources like shipping traffic or marine mammals. The ambient noise is mainly caused by agitation of the sea surface\,\cite{urick2}, i.e. by wind, breaking waves, spray, and cavitations. Thus it is correlated to the weather conditions, mainly to the wind speed. The model used for the simulation of the ambient noise is based on the so-called Knudsen spectra\,\cite{knudsen1948underwater}, which are adapted to the deep sea by applying attenuation effects. The power spectrum density (PSD) of the ambient noise is shown in Fig.~\ref{noise_psd}; the scatter plot presents the PSD of ambient noise for different levels of the wind speed and shipping traffic. The wind speed distribution included in the simulation corresponds to measurements at several weather stations near the coast of Marseilles, France. The mean noise level $\langle\sigma_{noise}\rangle$ is about 25\,mPa for the frequency range from 1\--100\,kHz and for 95\,\% of time the noise level is smaller than $2\langle\sigma_{noise}\rangle$\,\cite{Lahmann_Habil}. This is reproduced by the simulation as shown in Fig.~\ref{noise_occurrence}.\\
The simulation of the DAQ hardware comprises two parts: the simulation of the sensor and the read-out electronics response. The design of the DAQ hardware is inspired by the AMADEUS\,\cite{collaboration:2010fj} project. This includes acoustic sensors using the piezo-electric effect (hydrophones) and read-out electronics to amplify and digitise the signal. The inherent noise and the system transfer function, for both the sensors and the electronics, have been measured in the laboratory. Sensors normally show a directional dependency of their sensitivity, therefor signal and ambient noise have to be treated separately. In this case, the incident direction of the noise is the sea surface above the detector. Signal and ambient noise are then superimposed. The inherent noise of the sensor is added. For the read-out electronics, the resulting waveform is convoluted with the system transfer function and the inherent noise is added. The output is directed to a simulation of an on-line filter system\,\cite{Neff2009S185}, which, for real data, is used to reduce the amount of data to store and to pre-select for further off-line analysis. The filter is based on a matched filter, which uses a pre-defined bipolar pulse as reference to select signals with bipolar shape. In addition, a coincidence test between the sensors is performed.
\section{Applications}
\label{sec3:apps}
\begin{figure}[tb]
\centering
\includegraphics[width=0.4\textwidth]{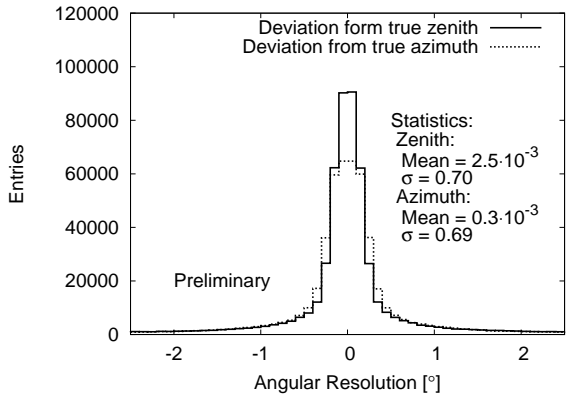}
\caption{
Angular resolution of the direction reconstruction algorithm shown for the zenith and azimuth angle. The mean of the distribution is around zero and the sigma is about 0.7$\,^{\circ}$ for both zenith and azimuth angle. The acoustic sources were generated in cube of 5\,$\times$\,5\,$\times$\,2.5\,km$^3$ around the detector centre.
\label{direction}}
\end{figure}

\begin{figure}[tb]
\centering
\includegraphics[width=0.4\textwidth]{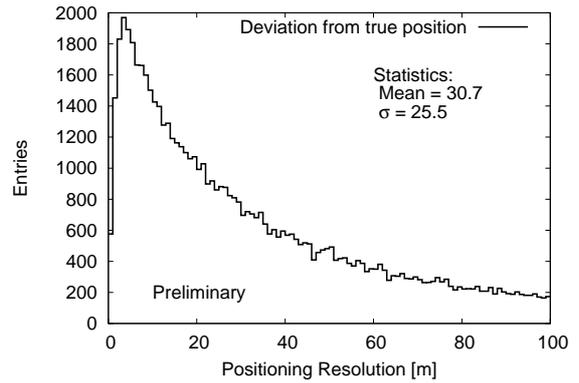}
\caption{
Resolution of the position reconstruction of the acoustic source shown as the distance between the reconstructed position and the true one. The resolution peaks at about 5\,m, the distribution is broad with a mean of about 30\,m and a sigma of about 25\,m for the given range. The acoustic sources for this analysis were generated in cube of 5\,$\times$\,5\,$\times$\,2.5\,km$^3$around the detector centre.
\label{position}}
\end{figure}

The simulation chain is used to study signal classification\,\cite{Neff:2011pd} and reconstruction algorithms\,\cite{Richardt:2009fr}. For the presented studies, a detector geometry similar to the configuration of AMADEUS\,\cite{collaboration:2010fj} is used, assuming six clusters of six sensors each at fixed positions and with fixed orientation. The precise reconstruction of the arrival time of the signal is crucial for the direction and position reconstruction of the acoustic source. The arrival time is determined by performing up-sampling of the filtered waveform sample and cross-correlation with a pre-defined bipolar pulse. This procedure achieves a precision of about $1\,\mu s$. Due to the narrow opening angle of the acoustic emission of a neutrino interaction, local clusters of sensors could be preferred in a detector design. Such clusters, consisting of several sensors arranged with interspaces of a few meters, have also advantages for the coincidence test used by the on-line filter and for the reconstruction of acoustic source position. The direction reconstruction is based on a least square fit of the measured arrival times at a given sensor cluster:
\begin{equation}
min ( \sum_i (t_{measured_{i}} - t_{expected_{i}}(\theta,\phi))^2 ),
\end{equation}
where $i \in 1..N$ ($N$ sensors of the cluster), $t$ the arrival time and $\theta$ and $\phi$ the zenith and azimuth angle, respectively. The acoustic sources for this analysis were generated in cube of 5\,$\times$\,5\,$\times$\,2.5\,km$^3$ around the detector centre. The angular resolution reached with this algorithm is centred around zero with a sigma of about 0.7$\,^{\circ}$ for both zenith and azimuth angle as shown in Fig.~\ref{direction}. This is consistent with the resolution of the reconstruction of the arrival times.
The position reconstruction of the acoustic source is obtained from triangulation of the previous determined direction of the incoming signal. If the directions were reconstructed for at least two of the sensor clusters, the triangulation is performed by minimising the distance between the rays starting at the sensor clusters and pointing into the reconstructed direction. The distribution of the distance between the reconstructed position and the true one peaks at about 5\,m, but is also broad with a mean of about 30\,m and a sigma of 25\,m (cf.~Fig.~\ref{position}). This is in agreement with the resolution of the direction reconstruction.\\
As a further application, signal classification will now be presented. The classification system stems from machine learning algorithms trained and tested with data from the simulation. As input, either an extracted feature vector or the filtered waveform is used; as output, binary class labels \--- bipolar or not \--- are predicted\,\cite{Neff:2011pd}. The Random Forest and Boosted Trees algorithms\,\cite{opencv} have achieved the best results for individual sensors and clusters of sensors. A Random Forest is a collection of decision trees. The classification works as follows: The Random Forest takes the input feature vector, makes a prediction with every tree in the forest, and outputs the class label that received the majority of votes. The trees in the forest are trained with different subsets of the original training data. Boosted Trees combine the performance of many so-called weak classifiers to produce a powerful classification scheme. A weak classifier is only required to be better than a random decision. Many of them smartly combined, however, result in a strong classifier. Decision trees are used as weak classifiers in this boosting scheme. In contrast to a Random Forest, the decision trees are not necessarily full-grown trees. For individual sensors, the classification error is of the order of 10\,\% for a well trained model. The combined results of the individual sensors in a cluster are used as new input for training. This method obtains a classification error below 2\,\%.
\section{Conclusion and Outlook}
\label{sec3:c&o}
As shown, the simulation chain is capable of reproducing the aspects necessary for acoustic neutrino detection \--- from the generation of the acoustic signal to different detector geometries and components. Also the deep-sea environment with its variable and diverse noise conditions is well replicated. Not yet integrated into the simulation is the refraction due to the depth dependence of the speed of sound in water. Furthermore, the whole processing chain for data can be tested as the simulation chain employs reconstruction and classification algorithms. The algorithms presented are the result of various studies to find the most suited ones. Detailed studies about the efficiency of on-line filters, effective volume and sensitivity of future large-scale detectors are currently pursued.
\section*{Acknowledgement}
The author would like to thank Sean Danaher for the provided source code. This work is supported by the german government (BMBF) with grants 05A08WE1 and 05A11WE1.
%
%
%
%
%
\bibliographystyle{model1-num-names}
\bibliography{biblio.bib}
%
%
%
\end{document}